\documentclass[aps,prl,superscriptaddress,twocolumn,showpacs,floatfix,a4paper]{revtex4-1}
\usepackage{amsfonts,amsmath,amssymb}
\usepackage{textcomp}
\usepackage{bm}
\usepackage{upgreek}
\usepackage{graphicx}
\usepackage{natbib}
\usepackage{color}
 \usepackage{times}
  \usepackage{bm}

\renewcommand{\vec}[1]{\bm{#1}}
\newcommand{\ee}{\mathrm{e}}
\newcommand{\ii}{\mathrm{i}}

\newcommand{\nablabf}{\boldsymbol{\nabla}}

\newcommand{\rot}{\nablabf\times}

\newcommand{\AAA}{\vec{A}}

\newcommand{\aaa}{\vec{a}}

\newcommand{\BBB}{\vec{B}}

\newcommand{\GGn}{\vec{G}}

\newcommand{\HHH}{\vec{H}}

\newcommand{\kkk}{\vec{k}}

\newcommand{\ppp}{\vec{p}}

\newcommand{\RRR}{\vec{R}}

\newcommand{\bra}[1]{\langle #1 |}
\newcommand{\ket}[1]{|#1\rangle}

\newcommand{\beq}[1]{\begin{equation} \eqlab{#1}}
	\newcommand{\eeq}{\end{equation}}
\newcommand{\bsub}{\begin{subequations}}
	\newcommand{\esub}{\end{subequations}}
\def\bal#1\eal{\begin{align}#1\end{align}}
\def\bsubal#1\esubal{\bsub \begin{align}#1\end{align} \esub}

\newcommand{\eqlab}[1]{\label{eq:#1}}
\renewcommand{\eqref}[1]{Eq.~(\ref{eq:#1})}

\newcommand{\figref}[1]{Fig.~\ref{fig:#1}}

\begin{document}
   	
\title{Graphene Nanobubbles as Valley Filters and Beamsplitters }	

\author{Mikkel Settnes}\email[]{mikse@nanotech.dtu.dk}
\affiliation{Center for Nanostructured Graphene (CNG), Department of Micro- and Nanotechnology Engineering, Technical University of Denmark, DK-2800 Kgs. Lyngby, Denmark}
\affiliation{Department of Photonics Engineering, Technical University of Denmark, DK-2800 Kgs. Lyngby, Denmark}
\author{Stephen R. Power}
\affiliation{Center for Nanostructured Graphene (CNG), Department of Micro- and Nanotechnology Engineering, Technical University of Denmark, DK-2800 Kgs. Lyngby, Denmark}
\affiliation{Department of Physics and Nanotechnology, Aalborg University, DK-9220 Aalborg, Denmark}
\author{Mads Brandbyge}
\affiliation{Center for Nanostructured Graphene (CNG), Department of Micro- and Nanotechnology Engineering, Technical University of Denmark, DK-2800 Kgs. Lyngby, Denmark}
\author{Antti-Pekka Jauho}
\affiliation{Center for Nanostructured Graphene (CNG), Department of Micro- and Nanotechnology Engineering, Technical University of Denmark, DK-2800 Kgs. Lyngby, Denmark}
%\author
%{Mikkel Settnes,$^{1-2}$ Stephen R. Power,$^{1,3}$  Mads Brandbyge,$^{1}$ Antti-Pekka Jauho$^{1\ast}$\\
%	\\
%	\normalsize{$^{1}$ Center for Nanostructured Graphene (CNG), DTU Nanotech, $^{2}$ DTU Fotonik} \\
%	 \normalsize{Technical University of Denmark, DK-2800 Kongens Lyngby, Denmark,}\\
%	%\normalsize{$^{2}$Department of photonics engineering, Technical University of Denmark,}\\
%	%\normalsize{ DK-2800 Kongens Lyngby, Denmark}\\
%	\normalsize{$^{3}$ Department of Physics and Nanotechnology,} \\
%	\normalsize{ Aalborg University, DK-9220 Aalborg, Denmark}
%	\\
%	\normalsize{$^\ast$To whom correspondence should be addressed; E-mail:  Antti-Pekka.Jauho@nanotech.dtu.dk}
%}
\date{\today}
%\author{Mikkel Settnes}
%\affiliation{Center for Nanostructured Graphene (CNG), DTU Nanotech, Technical University of Denmark, DK-2800 Kongens Lyngby, Denmark}
%\alsoaffiliation{Department of photonics engineering, Technical University of Denmark, DK-2800 Kongens Lyngby, Denmark}
%
%\author{Stephen R. Power}
%\affiliation{Center for Nanostructured Graphene (CNG), DTU Nanotech, Technical University of Denmark, DK-2800 Kongens Lyngby, Denmark}
%\alsoaffiliation{Department of Physics and Nanotechnology, Aalborg University, DK-9220 Aalborg Øst, Denmark}
%
%\author{Mads Brandbyge}
%\author{Antti-Pekka Jauho}
%\email{Antti-Pekka.Jauho@nanotech.dtu.dk}
%\affiliation{Center for Nanostructured Graphene (CNG), DTU Nanotech, Technical University of Denmark, DK-2800 Kongens Lyngby, Denmark}
% 	
%
%
% \keywords{Graphene, strain, pseudomagnetic field, quantum transport, valleytronics}

%%%%%%%%%%%%%%%%%%%%%%%%%%%%%%%%%%%%%%%%%%%%%%%%%%
%
% MAIN TEXT
%
%%%%%%%%%%%%%%%%%%%%%%%%%%%%%%%%%%%%%%%%%%%%%%%%%

%\baselineskip24pt

% Make the title.

\begin{abstract}
The low energy band structure of graphene has two inequivalent valleys at K and K$'$ points of  the Brillouin zone.
The possibility to manipulate this valley degree of freedom defines the field of valleytronics, the valley analogue of spintronics.
A key requirement for  valleytronic devices is the ability to break the valley degeneracy by filtering and spatially splitting valleys to generate valley polarized currents.
Here we suggest a way to obtain valley polarization using strain-induced inhomogeneous pseudomagnetic fields (PMF) which act differently on the two valleys. Notably, the suggested method does not involve external magnetic fields, or magnetic materials, as in previous proposals.
 In our proposal the strain is due to experimentally feasible  nanobubbles, whose associated PMFs lead to different real space trajectories for K and K$'$ electrons, thus allowing the two valleys to be addressed individually.
In this way, graphene nanobubbles can be exploited in both valley filtering and valley splitting devices, and our simulations reveal that a number of different functionalities are possible depending on the deformation field.
\end{abstract}
\maketitle

 	%\maketitle
 	%\date{\today}
%	\begin{abstract}	
%The low energy band structure of graphene presents two inequivalent valleys K and K$'$ in the Brillouin zone.
%The possibility to manipulate this valley degree of freedom motivates the field of valleytronics, the valley analogue of spintronics.
%A key objective is the ability to break valley degeneracy by filtering and spatially splitting valleys to allow the generation of valley polarized currents.
%Here we suggest a way to obtain valley polarization using only a strain-induced pseudomagnetic field (PMF) acting differently on the two valleys.
%We demonstrate that experimentally feasible strained nanobubbles and their associated PMFs lead to different real space trajectories for K and K$'$ electrons, allowing the two valleys to be addressed individually.
%In this way, graphene nanobubbles can be exploited in both valley filtering and valley splitting devices without the need for additional fields or interactions.
% 	\end{abstract}

%  \section*{Introduction}
A remarkable feature of Dirac fermions in graphene is the unique coupling between mechanical deformation and electronic structure.
Deforming the graphene lattice introduces an effective gauge field $\AAA$ in the low energy Dirac spectrum\cite{Guinea2010,Juan2011}, causing a pronounced sublattice polarization \cite{Neek-Amal2013,Schneider2015,Settnes2016}.
One can associate  a pseudomagnetic field (PMF) with this gauge field, $\BBB_s = \rot \AAA$. 
%\textcolor{red}{The gauge field, $\AAA \propto \big( \eps_{xx}-\eps_{yy},-2\eps_{xy} \big)$, %\cite{Vozmediano2010,Ramezani2013} is related to the strain tensor given within elasticity theory\cite{LandauBook} by %$\eps_{ij}(x,y) = \big[\pp_j u_i +\pp_i u_j + (\pp_i z) (\pp_j z)\big]/2$ with $\uuu$ and $z$ being the in-plane and %out-of-plane displacements, respectively. CAN WE DELETE?}
The presence of constant PMFs in graphene has been %studied theoretically \cite{Neek-Amal2013,Guinea2010,Settnes2016} and
spectacularly illustrated by scanning tunneling experiments, revealing signatures of Landau quantization \cite{Levy2010,Lu2012,Bai2015,Lim2015}. In contrast to the usual constant PMFs, in this Letter we focus on spatially varying PMFs, and show that inhomogenous PMFs can be used as a building block for valleytronic devices.

Unlike real magnetic fields, strain-induced PMFs conserve time-reversal symmetry and take opposite signs in the K and K$'$ valleys \cite{Vozmediano2010,Pereira2009prl}.
The effective gauge field enters the low energy Dirac Hamiltonian $H = v_F \bm{\sigma} \cdot \ppp$ via the transformation $\ppp \rightarrow \ppp \pm e \AAA$, where $\pm$ denote either the K or K$'$ valley \cite{Vozmediano2010,Pereira2009prl}.
This sign difference between K and K$'$, together with the spatially varying PMF, lies at the heart of our suggestion to manipulate the valley degree of freedom using strain engineering.
%The gauge field is given by $\AAA = -\hbar \beta \big( \eps_{xx}-\eps_{yy},-2\eps_{xy} \big)/2e a_0$, where $\beta\approx 3.37$, $e$ is the electron charge and $a_0=0.142$ nm is the C-C distance in the graphene lattice\cite{Vozmediano2010}. Further, the strain tensor is $\eps_{ij}(x,y) = \big[\pp_j u_i +\pp_i u_j + (\pp_i z) (\pp_j z)\big]/2$ with $\uuu$ and $z$ being the in-plane and out-of-plane displacements.
%The strain tensor $\eps_{ij}(x,y) = \big[\pp_j u_i +\pp_i u_j + (\pp_i z) (\pp_j z)\big]/2$ is given in terms of in-plane $\uuu$ and out-of-plane displacements $z$ and leads to the gauge field %$\AAA \sim \big( \eps_{xx} - \eps_{yy}, -2\eps_{xy}\big)$
%\bal
%\AAA = -\frac{\hbar \beta }{2 e a_0}\begin{pmatrix}
%	\eps_{xx} - \eps_{yy} \\ - 2 \eps_{xy}
%\end{pmatrix} \eqlab{Afield}
%\eal
%where $\beta\approx 3.37$, $e$ is the electron charge and $a_0=0.142$ nm is the C-C distance in the graphene lattice\cite{Vozmediano2010}.
%\eqref{Afield} only contains the leading order corrections to the tight binding representation of graphene including strain effects. Expansion to higher orders leads to effects like Fermi surface anisotropy and space dependent Fermi velocity\cite{deJuan2013,Pellegrino2011}.
 	 	
\begin{figure}[b!]
	\begin{center}
		\includegraphics[width= 0.42\textwidth]{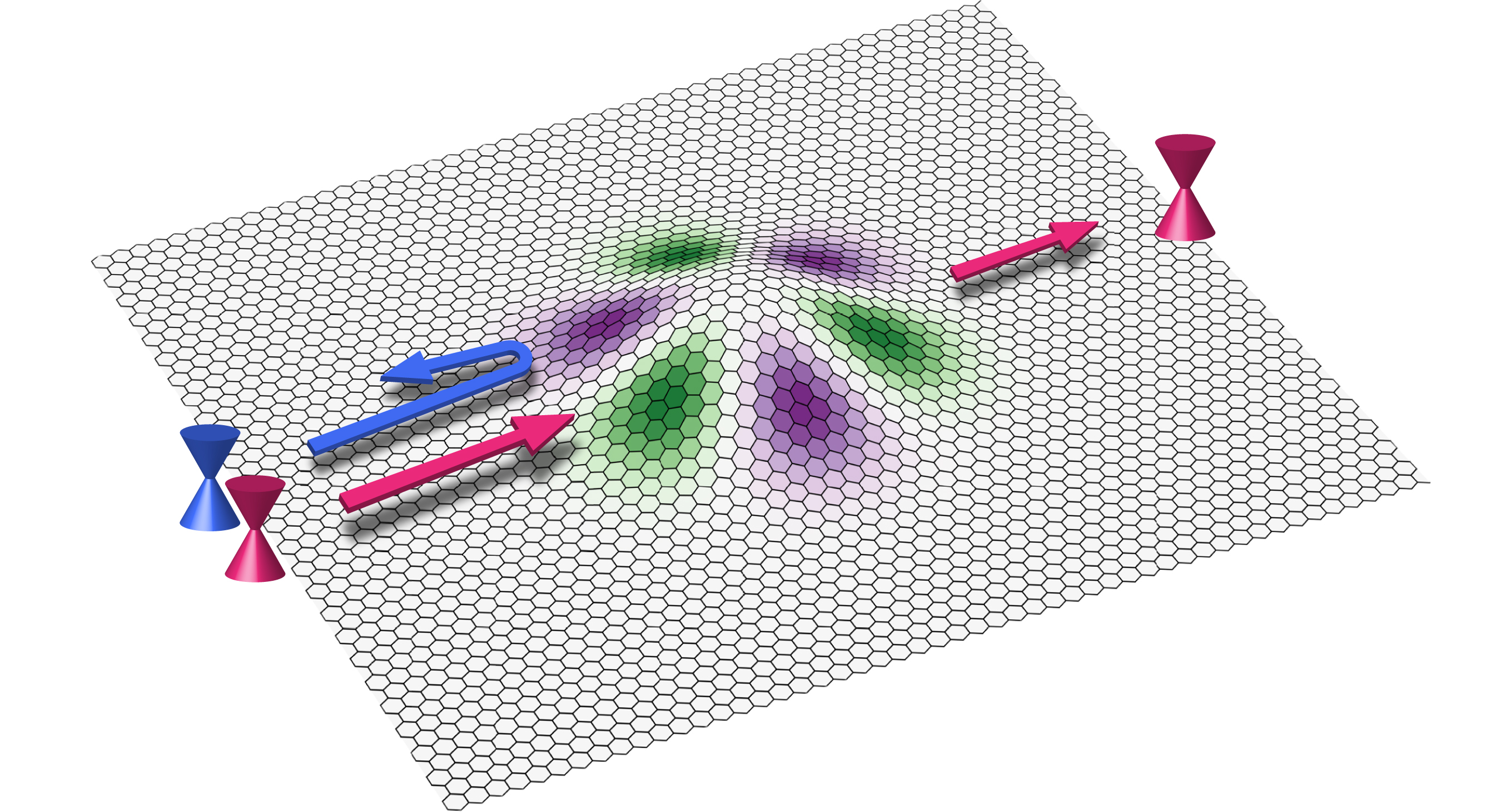}
		\caption[]{An incoming incoming electron wave containing both K and K$'$ valleys incident on a Gaussian nanobubble experiences the associated pseudomagnetic field indicated by the (green/purple) colormap. K valley electrons are backscattered whereas those from the K$'$ valley are transmitted due to the different trajectories imposed by the effective magnetic field for each valley when electrons are incident along a specific direction relative to the field.
		} \label{fig:system_sketch}
	\end{center}
\end{figure}
\begin{figure*}[htb]
	\begin{center}
		\includegraphics[width= 0.8\textwidth]{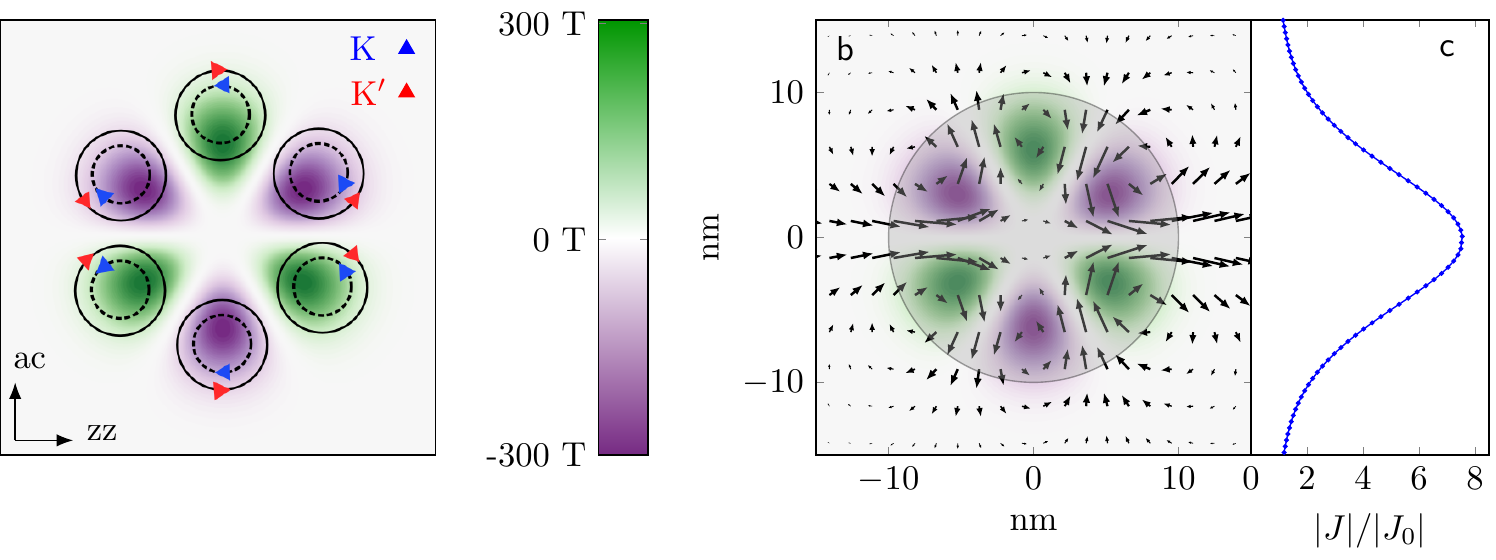}
		\caption[]{  %{\bf Pseudomagnetic field and current vortices caused by a Gaussian deformation. }
			(a) The colormap indicates the three-fold symmetric pseudomagnetic field caused by the circularly-symmetric Gaussian deformation. The vortices show trajectories corresponding to the field experienced in the K (blue) and K$'$ valleys (red). (b) Calculation of the local current incident from the left along the zigzag direction at $E=0.01|t_0|$.
			%, where $t_0$ is the hopping parameter between carbon atoms.
			The arrows indicate the direction and magnitude of the current. The arrows are averaged over several sites to enhance visibility. The shaded area indicates the $r < 2\sigma$ region. (c), Spatially resolved current density with strain ($|J|$), relative to that without strain ($|J_0|$), evaluated at the rightmost edge of (b), demonstrating that the strained region focuses the initially uniform current.	
		} \label{fig:gauss_1}
	\end{center}
\end{figure*}

Different routes have been suggested to create valley polarization \cite{Gunawan2006a,Gunawan2006b} in graphene\cite{Xiao2007_PRL,Gunlycke2011,Rycerz2007}, relying on nanoribbons/constrictions \cite{Rycerz2007,Chaves2010,Cavalcante2016,Carrillo-Bastos2016}, interplays between external fields\cite{Qi2013,Low2010,Fujita2010,daCosta2015}, spin-orbit coupling\cite{Grujic2014,Wu2016} or spatial/temporal combinations of gating and magnetic fields\cite{Zhenhua2011,Jiang2013,Cosma2014}. However, an experimental verification has proven to be challenging as practical and effective methods to manipulate the valleys in realistic setups  still need to be established.

In this Letter, we show that experimentally feasible local strain fields due to local deformations give rise to PMFs that allow for valley control without the need for additional fields or interactions. Experimental methods for producing such controllable strain fields include direct applied pressure from STM-tips\cite{Klimov2012}, gas-inflation\cite{Bunch2008,Georgiou2011,Zenan2014,Bahamon2015,Sloan2013} and substrate engineering\cite{Kusminskiy2011,Reserbat2014,Jones2014,Neek-Amal2012a,Neek-Amal2012b,Neek-Amal2012,Gill2015,Scharfenberg2011,Bao2009}.
Most of these approaches result in spatially localized strain fields taking the form of a pseudomagnetic dot.
%Nanobubbles are therefore highly interesting systems due to the controllable manner in which local PMFs of great magnitude can be induced in the graphene sheet.
The PMFs created this way are usually of great magnitude and local but spatially varying. We show that such systems can exhibit strong valley dependent effects associated with two key valleytronic components  -- namely valley filters (as illustrated in \figref{system_sketch}) and beam splitters, which spatially separate the different valleys (see \figref{inplane_1}).

\textit{Methodology:}
%The presented calculations envisage a spatially uniform electron wave impinging on the strained nanobubble, and compute its spatially and valley resolved transmission properties within the Landauer-B{\"u}ttiker framework using the recursive Patched Green's function approach \cite{Settnes2015,Power2011}.
The electronic structure of strained graphene is treated using a first nearest neighbor tight binding model $ \HHH=-\sum_{\langle i,j\rangle} t_{ij} c_i^\dagger c_j $,
where the sum $\langle i,j\rangle$ runs over nearest neighbors. Strain is included by modifying the hopping parameter such that $t_{ij} = t_0 \exp\big[-\beta (d_{ij}/a_0 -1)  \big]$ \cite{Pereira2009} where $a_0=0.142$ nm, $\beta = 3.37$, $t_0=-2.7$ eV and $d_{ij}$ is the modified bond length.
In this way, we do not use the Dirac model in the actual calculations but only to interpret the results from the full tight binding calculation.

We apply the Patched Green's function (PGF) approach \cite{Settnes2015,Power2011} to calculate the response of a plane electron wave impinging on the strained nanobubble. Using the PGF method, we replace the infinite graphene Hamiltonian by a finite effective Hamiltonian $\HHH_{\rm{eff}} = \HHH + \bm{\Sigma}_B$, where $\HHH$ describes a finite patch of the system and the self-energy $\bm{\Sigma}_B$ contains the influence of the surrounding infinite, pristine graphene sheet upon the patch.
The full Green's function for the patch region becomes
$
\GGn(E) = \big(E-\HHH-\bm{\Sigma}_B - \bm{\Sigma}_L\big)^{-1}, \eqlab{GFpatched}
$
where $\bm{\Sigma}_L$ is the lead self-energy describing a point-like metallic probe with a constant density-of-state emitting an electron wave with a mixture of both valleys \cite{Settnes2015}. The probe is placed 250 nm away from the deformations such that the impinging wave approximately becomes a plane wave.

$\bm{\Sigma}_B$ is expressed conveniently using pristine Green's functions along the boundary of the calculation area exploiting complex contour techniques \cite{Power2011,Settnes2014a}. To calculate the GF and the local current  $J_{ij} = \mathrm{Im} \big( t_{ij} A_{ij}\big)/\hbar$, we employ an adaptive recursive routine \cite{Settnes2015}.
Here the spectral function $A_{ij}$ is defined as ${A}_{ij} = \big(\bm{G} \bm{\Gamma}_L \bm{G}^\dagger\big)_{ij}$, where the broadening due to lead is $\bm{\Gamma}_L=\ii (\bm{\Sigma}_L -\bm{\Sigma}_L^\dagger)$.
%\footnote{The coupling is calculated using\cite{Meunier1998,Amara2007}, $|k,\lambda \rangle \cos\big(\theta_i\big)$ where $w_i = \ee^{-ad_i^2}/\sum_{j} \ee^{-ad_j^2}$, $\theta_i $ is the angle between the tip apex and site $i$, $\lambda=0.85\AA$, $a=0.6 \AA^{-2}$. $t_0$ is a scaling factor which we set to $t_0=10t$.}

To determine the valley occupation of the electron wave, we consider the outgoing scattering state in real space given by the spectral function, $ {A}_{ij}$ \cite{Paulsson2007}. %defined on the tight binding basis.
We expand this scattering state in the basis of the pristine eigenstate of graphene, $\ket{\kkk,\lambda}$ ($\lambda$ is the band index) \footnote{The pristine eigenstate of the graphene lattice is \unexpanded{$\ket{\kkk,\lambda}
 = \frac{1}{\sqrt{2N} } \sum_{\RRR_i} \ee^{-\ii \kkk \RRR_i } \bigg[ \ket{\RRR_i,\bullet} + \lambda \ee^{-\ii \phi_{\kkk}} \ket{\RRR_i,\circ} \bigg]
$}.
where $\RRR_i$ is the position of the unit cell, $\lambda=\pm$ is the band index and $\bullet$/$\circ$ denotes the sublattice A/B. Here,  $\ee^{-\ii \phi_{\kkk}} = f(\kkk)^*/|f(\kkk)|$
with $f(\kkk) = 1 + \exp(\ii \kkk \aaa_1) + \exp(\ii \kkk \aaa_2)$ and the lattice vectors $\aaa_{1} = a_0(-\sqrt{3},3)/2$ and $\aaa_{2} = a_0(\sqrt{3},3)/2$.}
In this way, the matrix element $c_{\kkk} = \bra{\kkk,\lambda}  \bm{A} \ket{\kkk,\lambda}$ becomes a spectral density in $\kkk$-space indicating the $\kkk$-values occupied by the real space scattering state. $c_{\kkk}$ can then be computed for each $\kkk$-value separately to produce a Fourier map of the scattering state illustrating the full valley occupation.
%We can use the Fourier map to define the occupation of one valley at a given point in space by summing $|c_{\kkk}|$ for $\kkk-\KKK < k_{\alpha}$, where $k_{\alpha}$ is a wavevector sufficiently large to contain the full signature around each $\KKK$/$\KKK'$-point without overlapping neighboring points.

%\section*{Results} 	
%\subsection*{Gaussian deformation for valley polarization}

\textit{Gaussian deformation for valley polarization:}
We first consider a Gaussian deformation \cite{Moldovan2013,Juan2011,Carrillo-Bastos2014} corresponding to an out-of-plane displacement
$
 z(r) = h_0 \mathrm{exp}\, (-r^2/2\sigma^2),
$
 where $h_0 = 3.5$ nm and $\sigma = 5$ nm are the height and width of the deformation, corresponding to a maximum strain of approximately 8.5\%. The result is robust and scalable for other deformation dimensions.
This circularly symmetric deformation gives rise to a PMF distribution indicated by the colormap in \figref{gauss_1}a for the K valley; an equally strong field but of opposite sign is experienced by the K$'$ valley.
The classical circular trajectories, forming a vortex pattern, expected for such a field profile are shown for the K (blue) and K$'$ valleys (red).
The plane wave is incident along the zigzag direction and the resulting local currents at an energy corresponding to the lowest resonance energy of the Gaussian deformation are shown in \figref{gauss_1}b.
A detailed description of the resonances is given in the Supplemental Material \cite{SM}.
The size and direction of the arrows in \figref{gauss_1}b indicate the magnitude and direction of the local current.
We especially note that the local current is largest at the interface between PMF regions of different sign, suggesting that snake states are formed here in a manner similar to systems with real magnetic fields\cite{Martino2007}.
From \figref{gauss_1}b-c it is also clear that the deformation enhances the current in the region directly behind it, acting as a lens which focuses the current at this electron energy \cite{Stegmann2016}.

Comparing the direction of the vortex patterns in \figref{gauss_1}a and the local current in \figref{gauss_1}b, we find that only the current direction associated with the K$'$ valley is visible.
This does not, however, imply that the bubble is in a valley-polarized eigenstate. Instead, only one of the trajectories matches the direction of the incoming wave.
Thus we find a pattern matching the K valley vortices for a current incident from the right, or a mixture of both patterns for incidence from top or bottom (see Supplemental Material for details\cite{SM}).
Electrons in the K valley ``see'' only the vortex pattern indicated by blue arrows in \figref{gauss_1}a which tends to backscatter electrons incident from left, and transmit through the dot if incident from the right.
Conversely, K$'$ electrons see the pattern shown by red arrows, and if incident from left they are guided through the strained region along snake states between regions with PMFs of opposite sign.
Thus, the valley selection mechanism relies on the symmetry breaking caused by the direction of the incoming current and not on a valley polarization of the states in the bubble.
The presence of states from the opposite valley of course makes the effects discussed here vulnerable to intervalley scattering, such as that induced by short-ranged disorder.

\begin{figure}[t]
	\begin{center}
		\includegraphics[width= 0.90\columnwidth]{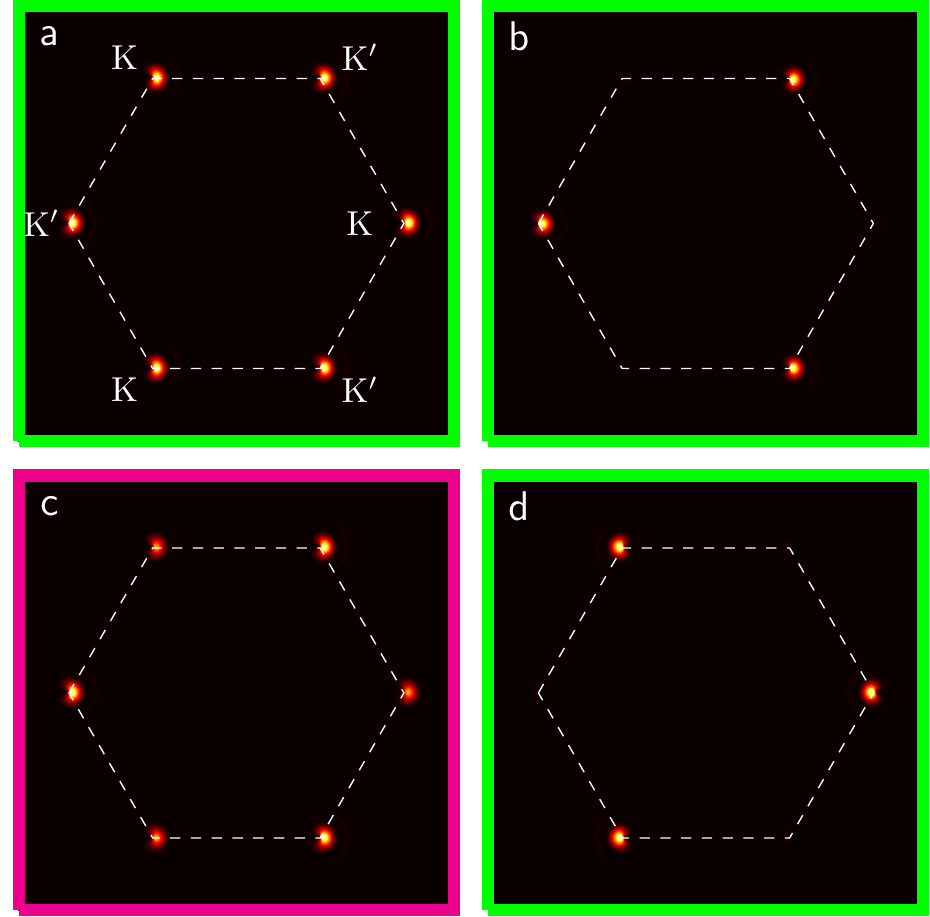}
		\includegraphics[width= 0.90\columnwidth]{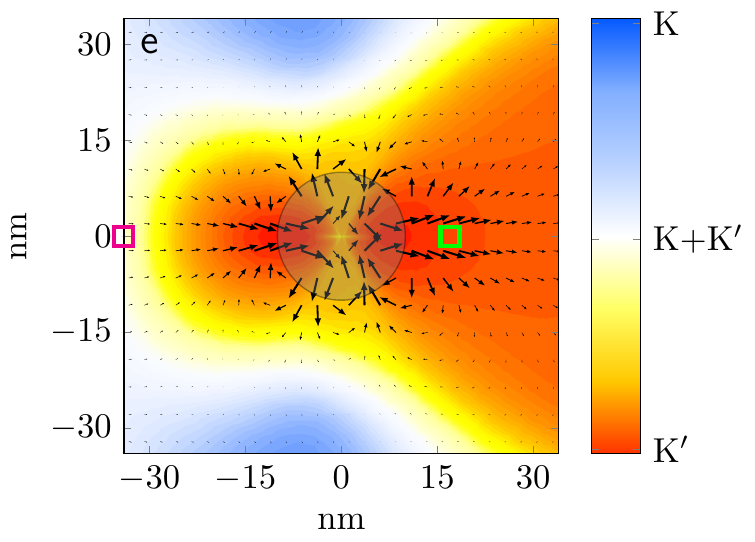}
		\caption[]{  %{\bf  The $k$-space occupation of the projected scattering state showing valley filtering for the Gaussian deformation. }
			%onto the pristine tight binding basis set
			(a) and (b) compare the $k$-space occupation for $E=0.01|t_0|$ at the green square in (e) without (a)  and with (b) the presence of the deformation. (c) $k$-space occupation for $E=0.01|t_0|$ at the red square in (e). (d), $k$-space occupation at the green square but with negative energy, $E=-0.01|t_0|$, showing the reversal of the $k$-filtering effect. (e), Real space map of the relative occupation of K and K$'$ in the scattering state showing the real space filtering of the valleys.
			The local current map from \figref{gauss_1}b is reproduced for convenience.		  }\label{fig:gauss_2}
	\end{center}
\end{figure}
 \begin{figure*}[t]
 	\begin{center}
 		\includegraphics[width= 0.75\textwidth]{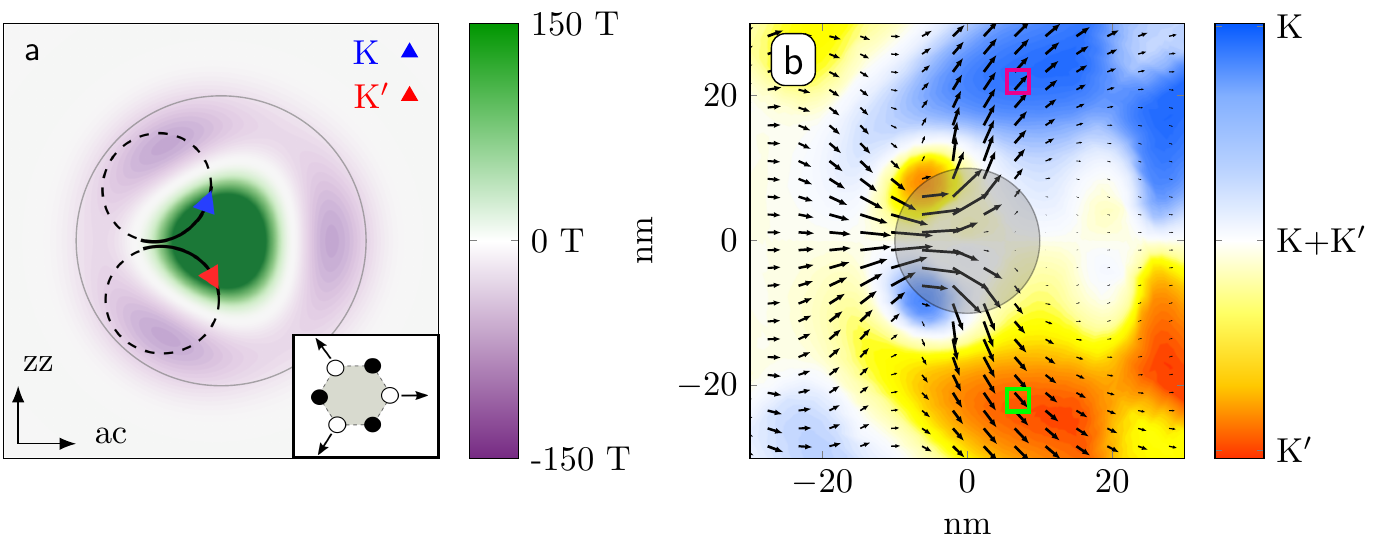}
 		\includegraphics[width= 0.15\textwidth]{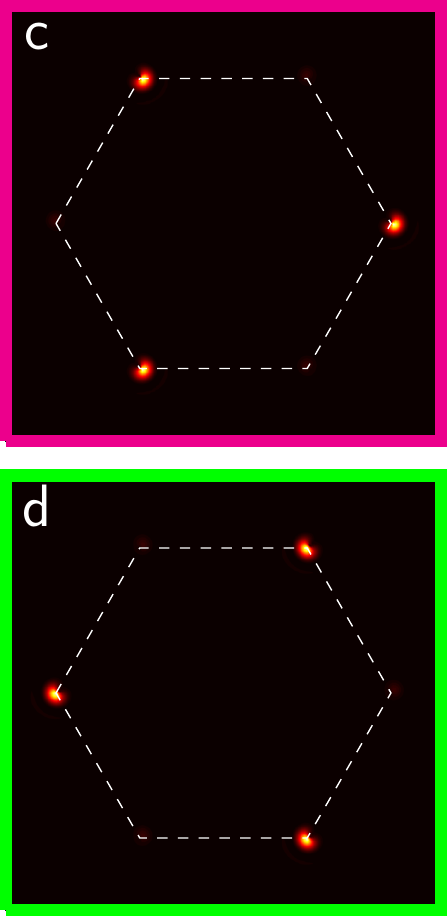}
 		\caption[]{ %\sffamily \small {\bf Spatial valley splitting caused by a triaxial deformation.}.
 			(a) Pseudomagnetic field distribution for the triaxial displacement field  with schematic trajectories shown for the K (blue) and K$'$ valley (red). The inset shows the direction of the triaxial strain. (b) Real space map at $E=0.019|t_0|$ of both the local current (arrows) and relative $k$-occupation (color map) of the scattering state incoming from the armchair direction (note the rotation compared to \figref{gauss_2}).
 			(c-d) Fourier maps for the scattering state at the red and green boxes indicated in {\bf b}.
 			%{\bf Valley splitting by triaxial deformation.}  (a) An incoming electron wave containing both K (blue) and K$'$ (red) valleys is split when incident on the triaxial deformation with the associated pseudomagnetic field indicated by the (green/purple) colormap. (b) K and K$'$ are deflected in different directions due to the different trajectories imposed by the effective magnetic field.
 		}\label{fig:inplane_1}
 	\end{center}
 \end{figure*} 	
%To further examine the valley dependence of the computed current patterns, we project the scattering state originating from the lead onto the pristine Bloch basis states of graphene, denoted $\ket{\kkk}$.
%The scattering states in a finite region can be generated from the appropriate real-space spectral function\cite{Paulsson2007}, ${A}^L$, with the required projection given by $c_{\kkk} = \bra{\kkk} {A}^L \ket{\kkk}$ (See Methods for more details).
%In this way, $c_{\kkk}$ becomes a spectral density in $k$-space indicating the $k$-values occupied by the real space scattering states at a given energy.
%Calculating $|c_{\kkk}|$ for all $k$-values in the Brillouin zone yields Brillouin zone maps showing the valleys present in the scattering states.
To further examine the valley dependence of the computed current patterns, we calculate the spectral density $|c_{\kkk}|$ for each $\kkk$-value.
Figs. \ref{fig:gauss_2}a-b show such Fourier maps generated for the region indicated by the green box in \figref{gauss_2}e without (\figref{gauss_2}a) and with (\figref{gauss_2}b) the presence of the strain field.
The valley filtering occurs when passing through the Gaussian deformation and the transmitted wave consists almost exclusively of electrons in the K$'$ valley.
On the other hand, \figref{gauss_2}c shows that both valleys are present (but not with equal weights) in the region before the bubble, shown by the red box.
Finally, the full map of the valley occupation (\figref{gauss_2}e) confirms the earlier intuitive analysis based on the local current trajectories in \figref{gauss_1}a: the K$'$ valley totally dominates the strained region while the K valley entirely avoids the strained region.
	
The valley filtering effect arises due to the different signs of the PMFs experienced by the two valleys. Similarly, the current paths for $E$ and $-E$ are equal, but the opposite energy sign swaps the valleys and the other valley is transmitted/backscattered (as illustrated in \figref{gauss_2}d).
This intriguing observation opens the desirable possibility of valley selectivity by a simple back gate as the Fermi energy is shifted between positive and negative values.
Even further tunability is possible when observing that the valley filtering effect is strongest at energies corresponding to resonances of the deformation.
Thus, small adjustments in the Fermi energy allow one to turn on and off the valley filtering effect.
In the Supplemental Material \cite{SM}, we examine how varying a gate allows one to move in and out of resonance with the eigenstates which are strongly affected by the PMF.

% \subsection*{Triaxial deformation for valley splitting}

\textit{Triaxial deformation for valley splitting:}
Finally, we consider an alternative geometry consisting of an in-plane triaxial strain \cite{Guinea2010,Juan2011,Levy2010} and additional out-of-plane deformation, appropriate for small bubbles formed by graphene on a substrate \cite{Levy2010}. The displacement are
\bal
\begin{pmatrix}
	u_r\\u_{\theta}\\ z
\end{pmatrix} = \begin{pmatrix}
u_0 r^2 \sin(3\theta) \\
u_0 r^2 \cos(3\theta)\\
h_0
\end{pmatrix} \ee^{-\frac{r^2}{2\sigma^2}} \,. \eqlab{inplane}
\eal
where $(r,\theta)$ are polar coordinates ($\theta=0$ corresponding to the zigzag direction). $h_0=1$ nm and $\sigma = 5$ nm are the height and width of the deformation and $u_0$ is the in-plane strength, which is chosen to give a PMF of approximately 300 T at the center of the deformation. For the chosen deformation size this yield a strain of approximately 2.5 \%.
The resulting PMF distribution is shown in \figref{inplane_1}a together with effective trajectories for the K (blue) and K$'$ valley (red) giving rise to a splitting of the current.
The real space valley polarization of the scattering state and the local currents are mapped in \figref{inplane_1}b for one resonant mode of the deformation together with the Fourier maps illustrating the valley splitting in \figref{inplane_1}c-d.
The valley-dependent electron trajectories are again governed by interfaces between regions with different PMF polarity which support propagation in opposite directions for the two valleys.
The net effect of such trajectories are a symmetric splitting of the valleys perpendicular to the incident (armchair) direction.
Details of non-symmetric incidence, higher order resonance modes and their resulting trajectories are given in the Supplemental Material \cite{SM}.

%\begin{figure}[t]
% 	\begin{center}
% 		%\includegraphics[width= 0.7\textwidth]{pics/BondCurrent/PMF_inplane_vortex1.pdf}
% 		\includegraphics[width= 0.45\textwidth]{alpha_inplane_quiver}\\
% 		\includegraphics[width= 0.4\textwidth]{inplane_kmap_panel_wide.pdf}
% 		%\includegraphics[width= 0.39\textwidth]{pics/BondCurrent/alpha_inplane_Quiver_ScatCross}
% 		%\includegraphics[width= 0.2\textwidth]{pics/Kmap/kmap_gauss_inplane_P3H_R50_H5_Px220_Py-70.pdf}
% 		%\includegraphics[width= 0.2\textwidth]{pics/Kmap/kmap_gauss_inplane_P3H_R50_H5_Px-220_Py-70.pdf}
% 		\caption[]{ {\bf $k$-space occupation of scattering state for triaxial deformation.}  (a) Real space map at $E=0.019|t_0|$ of both the local current (arrows) and relative $k$-occupation (color map) of the scattering state incoming from the armchair direction (note the rotation compared to \figref{gauss_2}).
% 			(b-c) Fourier maps for the scattering state at the red and green boxes indicated in (a).
% 		} \label{fig:inplane_2}
% 	\end{center}
%\end{figure}	

%\section*{Discussion}	    	
%In summary,
\textit{Discussion \& Conclusion:}
Direct experimental confirmation of the valley splitting in an experimental setting has previously been envisioned by employing real magnetic fields to alter the magnitudes, and not just the sign, of the total field experienced by each valley\cite{Zhenhua2011,Low2010}.
The results presented in this Letter, however, open a different route to experimental confirmation.
Using individually gated nanobubbles, we can exploit the interchanged roles of the valleys for opposite electron energies. In this way, oppositely gated nanobubbles will filter opposite valleys and effectively block the current while also creating the opportunity to turn on an off the valley polarized current.
Furthermore, the valley polarized currents generated by our setup will change the expected degeneracies of current-carrying states in, for example, Hall effect measurements.

To conclude, we have demonstrated how interfaces between pseudomagnetic fields of different polarity enable valley-dependent guiding of electrons in graphene. The two valleys experience a different field giving rise to different electron trajectories for each valley.
%In the case of locally strained bubble systems, the symmetry of the underlying strain field relative to the propagation direction of an incident electron wave determines how electrons from a given valley are guided through or around the deformation region.
The two nanobubble geometries considered provide illustrative examples of valley-filtering and valley-splitting devices,
 %where further tuning is possible by application of external magnetic fields or spatially-varying gates, or by combining multiple such bubbles in series,
 allowing for the construction of various valleytronic devices. This suggests alternative routes to experimental observation of valley polarization in graphene as well as
 %fundamental prospects for strain engineered graphene as
 a basis for topological valley (Hall) currents, along the lines of recent demonstrations by alternative mechanisms in graphene \cite{Gorbachev2014}, bilayer graphene \cite{Shimazaki2015} and other two dimensional materials \cite{Lensky2015,Mak2012,Lee2016}.

\textbf{Acknowledgements:} The authors thank N.A. Mortensen and T. Gunst for helpfull comments and discussions.  The work of M.S is supported by the Danish Council for Independent Research (DFF-5051-00011). The Center for Nanostructured Graphene (CNG) is sponsored by the Danish National Research Foundation (DNRF103).

\end{document}